\documentclass[onecolumn,amsmath,amssymb,nofootinbib,12pt]{article}
\usepackage{jheppub}
\usepackage{ifpdf}
\usepackage{graphicx,subcaption}
\usepackage{amsfonts}
\usepackage{amsmath}
\usepackage{amssymb}
\usepackage{epsfig}
\usepackage{pdfpages}
\usepackage{graphicx,epstopdf}
\usepackage[makeroom]{cancel}
\hypersetup{pdftex,colorlinks=true,allcolors=blue}
\usepackage{array}
\usepackage{ulem}

\def\be{\begin{equation}}
\def\ee{\end{equation}}
\def\ba{\begin{eqnarray}}
\def\ea{\end{eqnarray}}

\def\lf{\left}
\def\rt{\right}

\def\lf{\left}\def\rt{\right}                         
              \def\.{\cdot}

\usepackage{xspace}
\begin{document}
\title{\Large Subregion complexity in holographic thermalization with dS boundary}
\author[]{Shao-Jun Zhang}
\affiliation[]{Institute for Theoretical Physics \& Cosmology, Zhejiang University of Technology,
Hangzhou 310023, China}
\emailAdd{sjzhang84@hotmail.com}
\date{\today}
\abstract{We study the time evolution of holographic subregion complexity (HSC) in Vaidya spacetime with dS boundary. The subregion on the boundary is chosen to be a sphere within the cosmological horizon. It is found that the behaviour of HSC is similar to that in cases with flat boundary. The whole evolution can be divided into four stages: First, it grows almost linearly, then the growth slows down; After reaching a maximum it drops down quickly and gets to saturation finally. The linear growth rate in the first stage is found to depend almost only on the the mass parameter. As the subregion size approaches the cosmological horizon, this stage is expected to last forever with the subsequent three stages washed out. The saturation time $\tilde{t}_{sat}$ depends almost only on the subregion size $\tilde{R}$ as $\tilde{t}_{sat} = \tanh^{-1} (\tilde{R})$ which is linear in $\tilde{R}$ when $\tilde{R}$ is small but logarithmically divergent as $\tilde{R}$ approaches the cosmological horizon.}
\maketitle

\section{Introduction}

In past decades, with the idea of AdS/CFT or the more generic holographic principle~\cite{Maldacena:1997re,Gubser:1998bc,Witten:1998qj}, physicists are trying to build a bridge connecting gravity and other areas of modern theoretical physics, such as condensed matter physics (CMT)~\cite{Hartnoll:2009sz,Herzog:2009xv,McGreevy:2009xe,Horowitz:2010gk,Cai:2015cya}, QCD~\cite{Mateos:2007ay,Gubser:2009md,CasalderreySolana:2011us}, cosmology~\cite{Banks:2004eb}, quantum information theory (QIT)~\cite{Swingle:2009bg,Swingle:2012wq,Qi:2013caa} and etc. It is hoped that this bridge may help us get insights into both the strongly coupled problems in the quantum field theory (QFT) side as well as the origin of spacetime in the gravity side. After decades' efforts, several precise correspondences between the two sides are proposed. Recently, Susskind and his collaborators conjecture that complexity of the boundary QFT may be related to the interior geometry of black hole in the gravity side~\cite{Susskind:2014rva}. In QFT (or QIT), complexity of a target state is an important concept defined as the minimum number of unitary operators (or gates) needed to prepare the state starting from some reference state (for example the vacuum). So far, this conjecture has been refined into two concrete proposals, namely the CV (complexity=volume) and CA (complexity=action) conjectures. In the CV conjecture, complexity of a state living on a time slice $\Sigma$ of the boundary equals to the extremal volume of a codimension-one hypersurface ${\cal B}$ in the bulk ending on $\Sigma$ at the boundary~\cite{Stanford:2014jda}, that is
\begin{eqnarray}\label{CV}
C_V (\Sigma) =\max_{\partial {\cal B} = \Sigma} \left(\frac{{\rm Vol} ({\cal B})}{G_N R}\right),
\end{eqnarray}
where $G_N$ is the gravitational constant and $R$ is some typical length scale of the bulk geometry, for example the AdS radius or the horizon radius. While the CA conjecture states that complexity of a state equals to the on-shell gravitational action evaluated on the so-called Wheeler-DeWitt (WDW) patch of the bulk spacetime~\cite{Brown:2015bva,Brown:2015lvg}. Each conjecture has its own merits and demerits respectively~\cite{Hashimoto:2018bmb}. Inspired by these ideas, an amount of work are raised to study the holographic complexity for various gravity models to check these proposals~\cite{Momeni:2016ekm,Cai:2016xho,Brown:2016wib,Couch:2016exn,Yang:2016awy,Chapman:2016hwi,
	Carmi:2016wjl,Pan:2016ecg,Brown:2017jil,Kim:2017lrw,Cai:2017sjv,Alishahiha:2017hwg,Bakhshaei:2017qud,
	Tao:2017fsy,Guo:2017rul,Zangeneh:2017tub,Alishahiha:2017cuk,Abad:2017cgl,Reynolds:2017lwq,Hashimoto:2017fga,Nagasaki:2017kqe,Miao:2017quj,Ge:2017rak,
	Ghodrati:2017roz,Qaemmaqami:2017lzs,Carmi:2017jqz,Kim:2017qrq,Cottrell:2017ayj,Sebastiani:2017rxr,
	Moosa:2017yvt,HosseiniMansoori:2017tsm,Zhang:2017nth,Reynolds:2017jfs,Chapman:2018dem,Chapman:2018lsv,Khan:2018rzm,Caputa:2018kdj,Feng:2018sqm,Liu:2019smx,Jiang:2019qea,Jiang:2019pgc,Jiang:2018sqj,Jiang:2018gft}.

The above two conjectures are for the whole boundary system which both are then extended to be defined on subsystem respectively in Refs.~\cite{Alishahiha:2015rta} and~\cite{Carmi:2016wjl} later, and they are now called holographic subregion complexity (HSC). The subregion CV proposal is a natural extension of the well-known Hubney-Rangamani-Takayanagi (HRT) holographic entanglement entropy (HEE) conjecture~\cite{Ryu:2006bv,Hubeny:2007xt}. Namely, complexity of a subregion ${\cal A}$ of the boundary system equals to the volume of the extremal codimension-one hypersurface $\Gamma_{\cal A}$ enclosed by ${\cal A}$ and the corresponding Hubney-Rangamani-Takayanagi (HRT) surface $\gamma_{\cal A}$~\cite{Ryu:2006bv,Hubeny:2007xt}, that is
\begin{eqnarray}\label{subCV}
	{\cal C} (\gamma_{\cal A}) = \frac{{\rm Vol (\Gamma_{{\cal A}})}}{8 \pi G_{d+1} L},
\end{eqnarray}
where $L$ is the AdS radius. Later studies suggest that it should be dual to the fidelity susceptibility in QIT~\cite{Alishahiha:2015rta,MIyaji:2015mia}. While in the subregion CA proposal, complexity of subregion ${\cal A}$ is given by the on-shell gravitational action evaluated on the intersection region between WDW patch and the so-called entanglement wedge~\cite{Czech:2012bh,Headrick:2014cta}. Also, lots of work and effort have been devoted to understand the holographic subregion complexity~\cite{Caputa2017,Caputa2017b,Czech1706,subBenAmi2016,subRoy2017,
	subBanerjee2017,subBakhshaei2017,subSarkar2017,subZangeneh2017,subMomeni2017,subRoy2017b,subCarmi2017,Chen:2018mcc,Ageev:2018nye,Ghodrati:2018hss,Zhang:2018qnt,Alishahiha:2018lfv,Alishahiha:2018tep}.

On the other hand, in the so-called "holographic thermalization" topic, the AdS/CFT duality has been applied successfully to study the physics in non-equilibrium processes, especially the thermalization process of hot QCD matter which is strongly coupled and produced in heavy ion collisions at the Relativistic Heavy
Ion Collider~\cite{Gelis:2011xw,Iancu:2012xa,Muller:2011ra,CasalderreySolana:2011us}.
According to the AdS/CFT dictionary, the thermalization process in the
boundary QFT system is dual to
a black hole formation process in the bulk which can be modelled simply by a Vaidya-like metric. There are already lots of work on this topic and many interesting results are obtained. For more details on this topic, please refer to the review~\cite{Balasubramanian:2011ur} and references therein. 

Complexity in the holographic thermalization process is also studied to investigate its time evolution behaviours under thermal quench. In Refs.~\cite{Chapman:2018dem,Chapman:2018lsv}, by applying the CV and CA conjectures, it is found that the late time growth of holographic complexity in the Vaidya spacetime is the same as that found for an eternal black hole. In Ref.~\cite{Chen:2018mcc}, the time evolution of subregion complexity is studied in the process with the subregion CV conjecture. And the results show that the subregion complexity is not always a monotonically increasing function of time. Actually, it increases at early time, but after reaching a maximum it decreases quickly and gets to saturation finally. For other related work, please see Refs.~\cite{Ageev:2018nye,Ageev:2019fxn,Ling:2018xpc,Fan:2018xwf,Jiang:2018tlu}

However, it should be noted that the boundary QFTs considered in the above mentioned work are usually living on the flat Minkowski spacetimes. It would be interesting to generalize the discussions to more realistic situations where QFTs lives on curved spacetimes, which may hep us to understand the extremely hot and condensed physics such as in the very early universe. Several holographic models of the quantum field theory in curved spacetimes (QFTCS) have already been proposed in de
Sitter (dS) spacetime and other cosmological
backgrounds (please refer to the review \cite{Marolf:2013ioa} for details). Here we would like to mention the work done in Ref.~\cite{Marolf:2010tg}, where an interesting
holographic model was built to
relate the QFTs living on the dS
boundary to the bulk Einstein gravity. Employing
this model, the
thermalization process of QFTs in dS
spacetime is studied holographically in Ref.~\cite{Fischler:2013fba}. By applying the holographic entanglement
entropy as a probe, the whole
thermalization is found to be similar to
the flat boundary
case~\cite{Liu:2013iza,Liu:2013qca} and can be divided into a sequence of
processes. Moreover, the saturation time is found to depend almost only on the
entanglement sphere radius. When the radius is
small, the saturation time is almost a linear increasing function of the radius, as expected to coincide with the
result of the flat boundary case at this
time~\cite{Balasubramanian:2011ur}.
However, when the radius becomes larger and larger to approach the cosmological horizon, the
saturation time blows up logarithmically. Later, the study is extended to include the effect of higher-derivative terms, such as the Gauss-Bonnet
correction~\cite{Zhang:2014cga}. And it is found that increasing the Gauss-Bonnet coupling will shorten the saturation time. Please also refer to Refs. \cite{Fischler:2014ama,Fischler:2014tka,Nguyen:2017ggc} for other related work on AdS/CFT with dS boundary.

As the deep connection between holographic entanglement entropy (HEE) and holographic subregion CV (HSCV), it would be interesting to study the time evolution of subregion complexity in the thermalization process of the QFTs living on dS spacetime within the above model. It is natural to ask the following questions: How the existence of the cosmological horizon affects the behaviour of HSCV? Whether the time evolution behaviours of HSCV can be used to describe the the whole thermalization process? Is there any difference between behaviours of HSCV and HEE? The main goal of this work is trying to address these questions.

The work is organized as follows. In the next section, we will give a brief review of the holographic model of QFTs in dS spacetime proposed in Ref.~\cite{Marolf:2010tg}, including the Vaidya-like solution. Then in Sec. III, 
we study in detail the time evolution of HSCV in the thermalization process. Due to the complication of the equations needed to solve, we rely mainly on numerical calculations. The final section is devoted to discussions and summary.

\section{Gravity solutions with dS boundary}

In this section, following Refs.~\cite{Fischler:2013fba,Zhang:2014cga}, we will briefly review the bulk solutions in Einstein gravity with a foliation such that the boundary metric corresponds to a de Sitter spacetime. Three relevant solutions will be presented, including a vacuum AdS, a static AdS black hole and its Vaidya-like cousin. 

\subsection{Action}

We consider $(d+1)$-dimensional Einstein-Hilbert action as follows
\ba\label{Action}
S = \frac{1}{16 \pi G_N} \int d^{d+1} x \sqrt{-g} \lf(R - 2\Lambda\rt),
\ea
where $G_N$ is the Newton constant and $\Lambda$ negative cosmological constant. The action gives the following equations of motion
\ba\label{EoMs}
R_{\mu\nu} - \frac{1}{2} g_{\mu\nu} R + \Lambda g_{\mu\nu}=0.
\ea

For asymptotically AdS spacetime, the metric can be written in the Fefferman-Graham form~\cite{Fefferman:1985}
\ba\label{FGForm}
ds^2 = \frac{L^2}{z^2} \left(g_{\mu\nu} (z,x)dx^\mu dx^\nu + dz^2\right),
\ea
where $L$ is the AdS raduis related to the cosmological constant as $\Lambda = -\frac{d(d-1)}{2 L^2}$. The dual quantum field theory lives at the conformal boundary  $z=0$ with a metric $ds^2_\Sigma = g_{\mu\nu} (0,x)dx^\mu dx^\nu$. In this paper, we are interested in cases where the boundary metric $ds^2_\Sigma$ corresponds to a dS spacetime in certain coordinates.

\subsection{AdS vacuum solution} 

The equations of motion (\ref{EoMs}) admit an AdS vacuum solution as
\ba\label{Vacuum}
&&ds^2 = \frac{L^2}{z^2} \lf(-f(r) g(z) dt^2 + f^{-1}(r) g(z) dr^2 + r^2 g(z) d\Omega_{d-2}^2 + dz^2\rt),\nonumber\\
&&f(r) =1- H^2 r^2 ,\qquad g(z) = \lf(1-
\frac{H^2 z^2}{4}\rt)^2.
\ea
The conformal boundary locates at $z=0$ with conformally reduced metric
\ba\label{dS}
ds^2_\Sigma = -f(r) dt^2 + f^{-1} (r) dr^2 + r^2 d\Omega_{d-2}^2,
\ea
which is just the dS spacetime in the static patch with a cosmological horizon at $r=1/H$, where $H$ denotes the Hubble constant.

The AdS vacuum solution is dual to the vacuum state of the dual QFT  with the latter can be taken as the well-known Bunch-Davis or Euclidean vacuum. For a geodesic observer sitting at $r=0$, the Bunch-Davis vacuum appears to have temperature $T_{dS} = H/2\pi$ natural for the existence of the cosmological horizon.

\subsection{AdS black hole solution}

The equations of motion (\ref{EoMs}) also admit an AdS black hole solution with the dS boundary (\ref{dS})
\ba\label{BH}
&&ds^2 = \frac{L^2}{z^2} \lf(-h(z) dt^2 + \frac{dz^2}{h(z)}+ \frac{H^2 L^2}{f(r)^2}dr^2 + \frac{H^2 L^2}{f(r)} r^2 d\Omega_{d-2}^2\rt),\nonumber\\
&&h(z) = 1 - \frac{z^2}{L^2} - \frac{m z^d}{L^{2(d-1)}}.
\ea
The event horizon $z_+$ is given by the largest positive root of $h(z)$. The mass parameter $m$ can be written in terms of the horizon as
\ba
m = \frac{L^{2(d-1)}}{z_+^d} \lf(1-\frac{z_+^2}{L^2}\rt).
\ea
The Hawking temperature of the black hole is
\ba\label{Temperature}
T_H = \frac{L^2 d -(d-2) z_+^2}{4\pi L^2 z_+}.
\ea
It should be noted that the zero temperature limit of the black hole solution (\ref{BH}) is the not the solution with $m=0$ which is isometric to the AdS vacuum solution (\ref{Vacuum}). Actually, the zero temperature limit of the solution has the smallest horizon radius and most "negative" mass as
\ba
z_+^{\textrm ext} = L \sqrt{\frac{d}{d-2}},\qquad m^{\textrm ext} = - \frac{2 L^{2(d-1)}}{(d-2) (z_+^{\textrm ext})^d}
\ea
This means that when the mass is negative in the range $0> m > m^{\textrm ext}$, the black hole still has a regular horizon and reasonable thermodynamics. This is a typical behavior of topological black holes.

Holographically, the black hole solution is dual to the QFT on the static patch of dS spacetime at the temperature given by Eq.~(\ref{Temperature}). Note that this temperature does not have to be the same as the dS temperature $T_{dS}$. For more discussions on this point, please refer to Ref.~\cite{Fischler:2013fba}.

\subsection{Vaidya-like solution}

Our aim is to study the holographic thermalization process of the dual QFT under quench. This process can be simply described holographically by a Vaidya-like geometry in the bulk.

Going to the Eddington-Finskelstein coordinates and introducing a time-dependent mass parameter, from the black hole solution one can obtain its Vaidya-like cousin as
\ba\label{Vaidya}
&&ds^2 =\frac{L^2}{z^2} \lf(-h(v,z) dv^2 -2 dv dz+ \frac{H^2 L^2}{f(r)^2}dr^2 + \frac{H^2 L^2}{f(r)} r^2 d\Omega_{d-2}^2\rt),\nonumber\\
&&h(v,z) = 1 - \frac{z^2}{L^2} - \frac{m(v) z^d}{L^{2(d-1)}}.
\ea
External source should be introduced to maintain the equations of motion
\ba
R_{\mu\nu} - \frac{1}{2} g_{\mu\nu} R + \Lambda g_{\mu\nu} = 8\pi G_N T_{\mu\nu}^{\textrm ext},\nonumber\\
8\pi G_N T_{\mu\nu}^{\textrm ext} = \frac{(d-1)z^{d-1}}{2 L^{2(d-1)}} \frac{d m(v)}{dv} \delta_{\mu v} \delta_{\nu v},
\ea
which implies that the infalling shell is made of null dust. We take the form of the mass function as
\ba
m(v) = \frac{M}{2} \lf[1 + \tanh\lf(\frac{v}{v_0}\rt)\rt],
\ea
where $M>0$ is the total mass of the dust shell and $v_0$ of its thickness. Then the solution describes the collapsing of the null dust shell from the boundary into the bulk to form a black hole. At the QFT side, it corresponds to a sudden global injection of energy into the system and then let it evolve from the Bunch-Davis vacuum  to a thermal state with $T>T_{dS}$.

\section{Holographic entanglement entropy and subregion complexity}

In this section, by applying the holographic subregion CV (HSCV) (\ref{subCV}), we will study the time evolution of holographic subregion complexity in the thermalization process which is described by the Vaidya-like geometry holographically. 

On the boundary at time $\tilde{t}$, taking into account the symmetry of the Vaidya-like metric (\ref{Vaidya}), it is convenient to choose the subregion ${\cal A}$ to be a $(d-1)$-dimensional sphere centred at $\tilde{r} =0$ ($\tilde{r} \equiv H r$) with raduis $\tilde{R}$. According to the conjecture (\ref{subCV}), the holographic subregion complexity of ${\cal A}$ is given by the extreme volume of the codimension-one hypersurface $\Gamma_{\cal A}$ enclosed by ${\cal A}$ and its corresponding HRT surface $\gamma_{\cal A}$. So, to study the holographic subregion complexity, we should first find the HRT surface $\gamma_{\cal A}$ whose area gives the holographic entanglement entropy.

\subsection{Holographic entanglement entropy}

Considering the symmetry, the HRT surface $\gamma_{\cal A}$ in the bulk can be parameterized by functions $z(\tilde{r})$ and $v(\tilde{r})$, with the boundary conditions
\ba
z(\tilde{R}) = \epsilon, v(\tilde{R}) = \tilde{t},
\ea
where $\epsilon$ is an UV cutoff constant. At the tip of the HRT surface, taking into account the symmetry, we have
\ba
z'(0) = v'(0)=0,z(0) = z_\ast, v(0)=v_\ast,
\ea
where $(z_\ast, v_\ast)$ are two parameters labelling the location of the tip and the prime denotes derivative with respect to $\tilde{r}$. The induced metric on $\gamma_{\cal A}$ is
\ba
ds^2 = \frac{L^2}{z^2} \lf(\frac{L^2}{(1-\tilde{r}^2)^2} - h(v,z) v'^2 - 2 v' z'\rt) d\tilde{r}^2 + \frac{L^4}{z^2} \frac{\tilde{r}^2}{1-\tilde{r}^2} d\Omega_{d-2}^2.
\ea
The holographic entanglement entropy functional is given by the area of the HRT surface
\ba\label{HEE}
&&{\cal S} = \frac{L^{2d-3}}{4 G_N} \Omega_{d-2} \int_0^{\tilde{R}} \frac{d\tilde{r}}{z^{d-1}} Q P^{d-2},\\
&&Q \equiv \sqrt{\frac{L^2}{(1-\tilde{r}^2)^2} - h(v, z) v'^2 - 2 v' z'},\qquad P \equiv \frac{\tilde{r}}{\sqrt{1-\tilde{r}^2}}.\nonumber
\ea
To find the extreme value of this functional, we need to solve the two equations of motion, which can be obtained by varying the functional and are rather complicated
\ba
&&h^2 \left(\tilde{r}^2-1\right)^2 v'^3 \left((d-1) \left(\tilde{r}^2-1\right) \tilde{r} z'+(d-2) z\right)\nonumber\\
&&+v' \left(\left(\tilde{r}^2-1\right) z' \left(\left(\tilde{r}^2-1\right) \left(2 z' \left((d-1) \left(\tilde{r}^2-1\right) \tilde{r} z'+(d-2) z\right)-\tilde{r} \left(\tilde{r}^2-1\right) z z''\right)-(d-1) h L^2 \tilde{r}\right)\right.\nonumber\\
&&\qquad\left.+h L^2 z \left(-d+2 \tilde{r}^2+2\right)\right)\nonumber\\
&&+3 h \left(\tilde{r}^2-1\right)^2 v'^2 z' \left((d-1) \left(\tilde{r}^2-1\right) \tilde{r} z'+(d-2) z\right)-(d-1) L^2 \tilde{r} \left(\tilde{r}^2-1\right) z'^2\nonumber\\
&&+L^2 z \left(-d+2 \tilde{r}^2+2\right) z'+v'' \left(h L^2 \tilde{r} \left(\tilde{r}^2-1\right) z+\tilde{r} \left(\tilde{r}^2-1\right)^3 z z'^2\right)+L^2 \left(\tilde{r}^2-1\right) \tilde{r} z z''=0,\nonumber\\
\ea
\ba
&&(d-1) h^2 \tilde{r} \left(\tilde{r}^2-1\right)^4 v'^4-2 (d-1) h L^2 \tilde{r} \left(\tilde{r}^2-1\right)^2 v'^2\nonumber\\
&&+z' \left(3 (d-1) h \tilde{r} \left(\tilde{r}^2-1\right)^4 \left(v'\right)^3-3 (d-1) L^2 \tilde{r} \left(\tilde{r}^2-1\right)^2 v'\right.\nonumber\\
&&\qquad\left.+z \left(\tilde{r} \left(\tilde{r}^2-1\right)^4 v' v''-2 (d-2) \left(\tilde{r}^2-1\right)^3 v'^2\right)\right)\nonumber\\
&&+z \left((2-d) h \left(\tilde{r}^2-1\right)^3 v'^3+L^2 \left(\tilde{r}^2-1\right) \left(d-2 \tilde{r}^2-2\right) v'-L^2 \tilde{r} \left(\tilde{r}^2-1\right)^2 v''\right)\nonumber\\
&&+(d-1) L^4 \tilde{r}+2 (d-1) \tilde{r} \left(\tilde{r}^2-1\right)^4 v'^2 z'^2-\tilde{r} \left(\tilde{r}^2-1\right)^4 z v'^2 z''=0.\nonumber\\
\ea
To avoid symbol confusion, we denote the solution of the above two equations as $(v_0(\tilde{r}), z_0(\tilde{r}))$ which parameterize the HRT surface. The relation between $v$ and $z$ on the HRT surface, denoted as $v_0(z)$, can be obtained by eliminating the parameter $\tilde{r}$ from the two functions.

Generally, the HEE (\ref{HEE}) is ultra-divergent. To remove the divergence and for convenience, we define a normalised HEE as
\ba\label{RHEE}
\hat{\cal S}  \equiv \frac{4 G_N ({\cal S}_{Vaidya} - {\cal S}_{AdS})}{V_{\cal A}},
\ea
where ${\cal S}_{AdS}$ is the HEE for the same subregion ${\cal A}$ in pure AdS geometry. And $V_{\cal A} \equiv L^{d-1} \Omega_{d-2} \int_0^{\tilde{R}} \frac{\tilde{r}^{d-2}}{(1-\tilde{r}^2)^{d/2}} d\tilde{r} = L^{d-1} \Omega_{d-2} \frac{\tilde{R}^{d-1}}{d-1} \ _2 F_1\left(\frac{d-1}{2},\frac{d}{2},\frac{d+1}{2}, \tilde{R}^2\right)$ is the volume of the subregion ${\cal A}$ \footnote{Actually, it should be noted that $V_{\cal A}$ is divergent as $\tilde{R}$ approaches the cosmological horizon to cover the whole boundary space.}. So, $\hat{\cal S}$ can be seen as a normalised entanglement entropy density.

\subsection{Holographic subregion complexity}

Due to the spherical symmetry, the co-dimension one extreme hypersurface $\Gamma_{\cal A}$, enclosed by ${\cal A}$ and the HRT surface $\gamma_{\cal A}$, can be parameterized by function $v = v(z, \tilde{r})$. The induced metric on $\Gamma_{\cal A}$ is
\ba
ds^2 = &&\frac{L^2}{z^2} \lf[-\lf(h \frac{\partial v}{\partial z} + 2\rt) \frac{\partial v}{\partial z} d z^2 - 2 \lf(h \frac{\partial v}{\partial z} + 1\rt) \frac{\partial v}{\partial \tilde{r}} dz d\tilde{r} \rt. \nonumber\\
&&+ \lf.\lf(\frac{L^2}{(1-\tilde{r}^2)^2} - h \lf(\frac{\partial v}{\partial \tilde{r}}\rt)^2 \rt) d\tilde{r}^2 + \frac{L^2 \tilde{r}^2}{1-\tilde{r}^2} d\Omega_{d-2}^2\rt].
\ea
According to the HSCV proposal (\ref{subCV}), the holographic subregion complexity functional of $\Gamma_{\cal A}$ is
\ba \label{HSC}
&&{\cal C} = \frac{L^{2d-2} \Omega_{d-2}}{8\pi G_N L} \int_0^{z_\ast} dz \int_0^{\tilde{r}(z)} d\tilde{r} N P^{d-2}{z^{-d}} ,\\
&&N \equiv \sqrt{-\lf[\frac{L^2}{(1-\tilde{r}^2)^2} - h v_{\tilde{r}}^2\rt](h v_z + 2) v_z - (h v_z + 1)^2 v_{\tilde{r}}^2},\qquad P \equiv \frac{\tilde{r}}{\sqrt{1-\tilde{r}^2}},\nonumber
\ea
where $v_z \equiv \frac{\partial v}{\partial z}$ and $v_{\tilde{r}} \equiv \frac{\partial v}{\partial \tilde{r}}$. To extremizing the HSCV functional, we need to solve the equation of motion which can be obtained by varying the functional with respect to $v(z, \tilde{r})$
\ba
&&L^2 \tilde{r} v_z \left(L^2 \left(v_z \left(2 d h \left(h v_z+3\right)-z v_z \left(h_v+h h_z\right)-3 z h_z\right)+4 d\right)-2 \left(\tilde{r}^2-1\right)^2 z v_{rr} \left(h v_z+2\right)\right)\nonumber\\
&&+2 L^2 \left(\tilde{r}^2-1\right) z v_r \left(\left(d-2 \left(\tilde{r}^2+1\right)\right) v_z \left(h v_z+2\right)+2 \left(\tilde{r}^2-1\right) \tilde{r} v_{zr} \left(h v_z+1\right)\right)\nonumber\\
&&+2 L^2 \left(\tilde{r}^2-1\right)^2 \tilde{r} v_r^2 \left(v_z \left(d h-z h_z\right)+d-h z v_{zz}\right)+2 (d-2) \left(\tilde{r}^2-1\right)^3 z v_r^3+2 L^4 \tilde{r} z v_{zz}=0.\nonumber\\
\ea
At first glance, it seems difficult to solve the above equation. However, it is interesting to note that $v(z,\tilde{r})=v_0(z)$ is just the solution (Here we would like to emphasize again that $v_0(z)$ is just the function giving the relation between $v$ and $z$ on the HRT surface), which can be checked directly by plugging $v_0(z)$ into the equation. This simply means that $\Gamma_{\cal A}$ is just formed by dragging the HRT surface $\gamma_{\cal A}$ along the $\tilde{r}$ direction. Similar feature has already been observed in flat boundary case with strip subregion in Ref.~\cite{Chen:2018mcc}.

As HEE, HSC is also ultra-divergent, so we can also define a normalised HSC density as
\ba
\hat{\cal C} \equiv \frac{8\pi G L ({\cal C}_{Vaidya} - {\cal C}_{AdS}) }{ V_{\cal A}},
\ea
where ${\cal C}_{AdS}$ is the HSCV for ${\cal A}$ in pure AdS geometry.

\subsection{Numerical results}

Having set up the general frame work of HEE and HSCV, now we are ready to study the time evolution of HSC in holographic thermalization. Due to the complication of the equations needed to solve, we rely on numerical method. And for convenience, we set the AdS radius $L=1$.

\subsubsection{General behaviours}

In Fig.~1, we plot the time evolution of normalised HSC density $\hat{\cal C}$ for various $\tilde{R}$ with fixed spacetime dimension. From the figure, one can see that the time evolution of $\hat{\cal C}$ is not a monotonically increasing function of the time. Rather, it can be divided into four stages: After quench, firstly it grows quickly and almost linearly, then the growth slows down; After reaching a maximal value $\hat{\cal C}_{max}$ it starts to drop down fast,and shortly after the drop down stops and it saturates to a constant value $\hat{\cal C}_{sat}$ finally. Moreover, it is interesting to note that the final saturation constant may be negative, which means that the final value of the complexity may be smaller than its initial value. These behaviours are very different from that in CV or CA conjectures, where the complexity is always a monotonically increasing function of time \cite{Chapman:2018dem,Chapman:2018lsv}. Similar behaviours have been observed in flat boundary cases with strip subregion \cite{Chen:2018mcc,Ling:2018xpc}, indicating universality of the behaviours.

\begin{figure}[!htbp]
	\centering
	\includegraphics[width=0.32\textwidth]{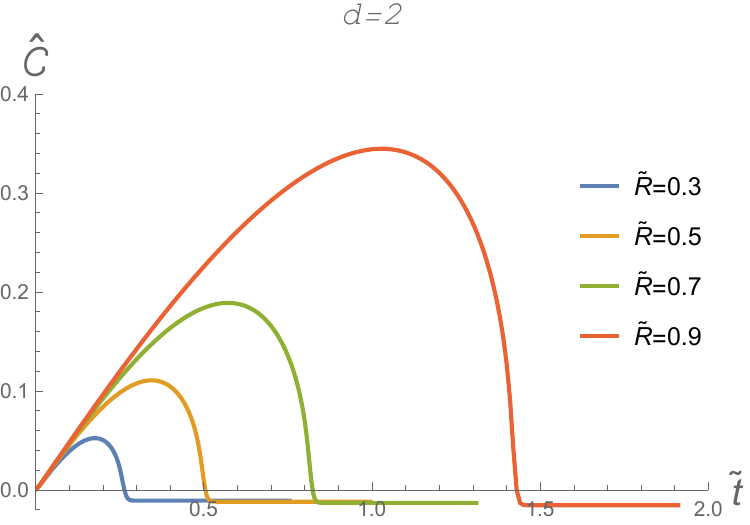}
	\includegraphics[width=0.33\textwidth]{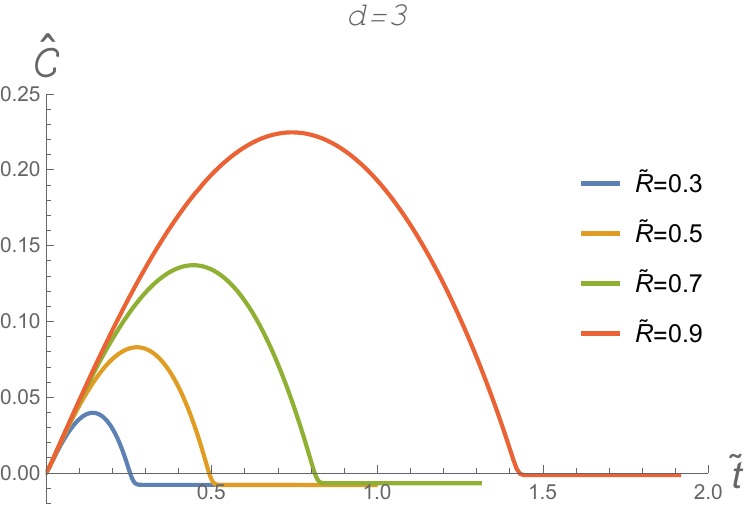}
	\includegraphics[width=0.32\textwidth]{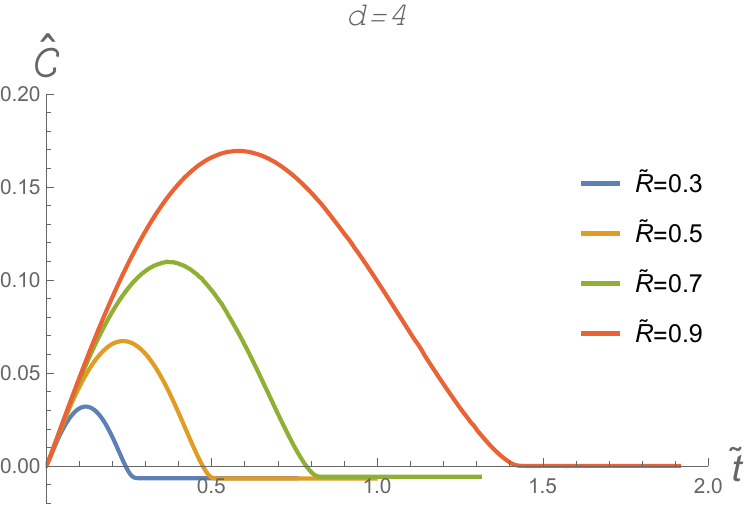}
\caption{(colour online) Time evolution of normalised HSCV density $\hat{\cal C}$ for various $\tilde{R}$ with fixed $d$. The mass is fixed as $M=1$.}
\end{figure}

\begin{figure}[!htbp]
	\centering
	\includegraphics[width=0.45\textwidth]{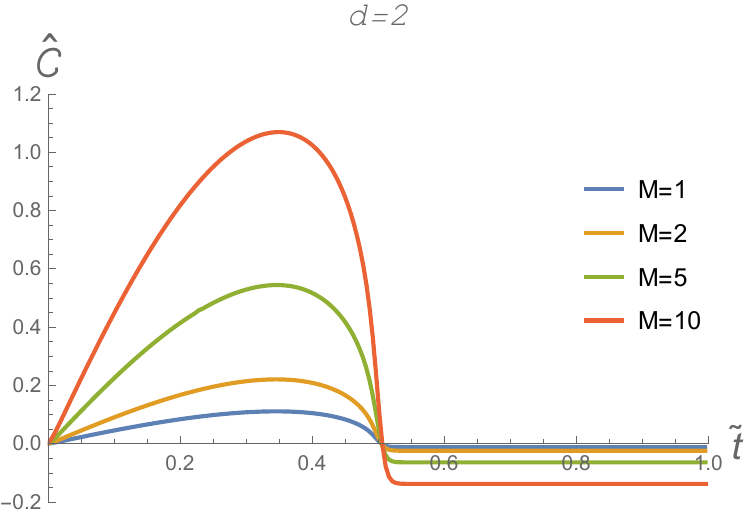}
	\includegraphics[width=0.45\textwidth]{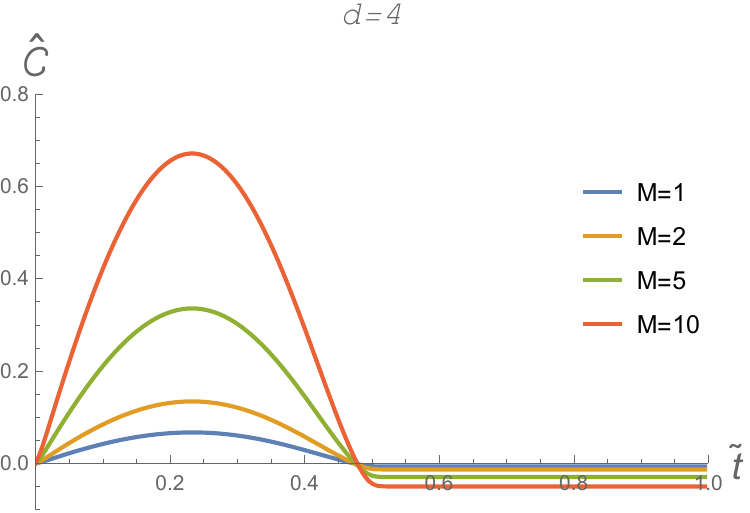}
	\caption{(colour online) Time evolution of normalised HSCV density $\hat{\cal C}$ for various $M$ with fixed $d$. The subregion size is fixed as $\tilde{R}=0.5$.}
\end{figure}

\begin{figure}[!htbp]
	\centering
	\includegraphics[width=0.45\textwidth]{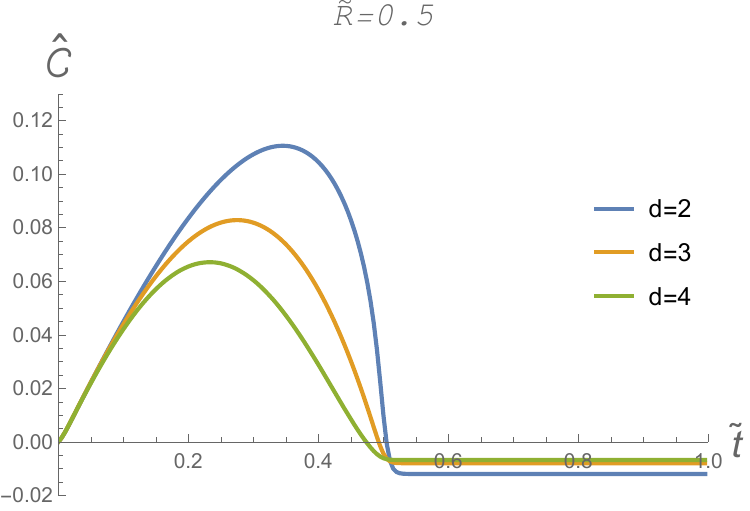}
	\includegraphics[width=0.45\textwidth]{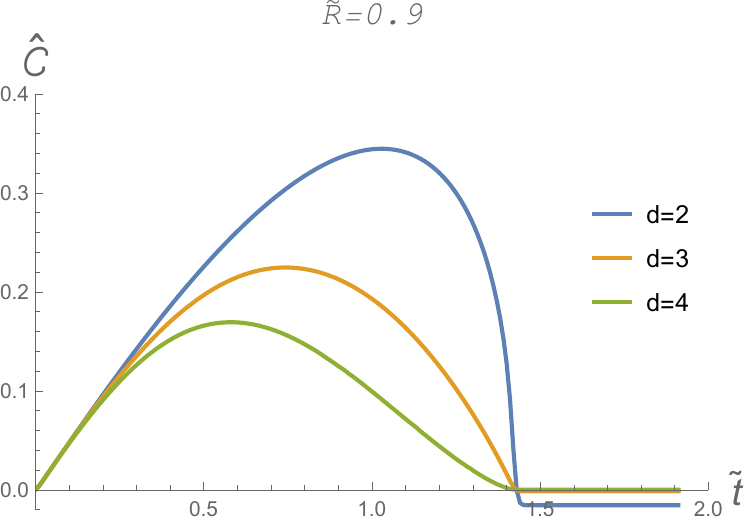}
\caption{(colour online) Time evolution of normalised HSCV density $\hat{\cal C}$ for various $d$ with fixed $\tilde{R}$. The mass is fixed as $M=1$. }
\end{figure}

From Fig.~1-3, one can see that the maximal value $\hat{\cal C}_{max}$ depends on the subregion size $\tilde{R}$, the spacetime dimension $(d+1)$ and the mass parameter $M$. Increasing $\tilde{R}$ or $M$ will yield a bigger $\hat{\cal C}_{max}$, while increasing the dimension will, on the contrary, lower the maximal value.

Moreover, one can also see that the final saturation constant $\hat{\cal C}_{sat}$ also depends on $(\tilde{R}, d, M)$ but in a more complicated way.

\subsubsection{Linear growth stage}

Let us focus on discussing the first stage when $\hat{\cal C}$ grows almost linearly in time, i.e.,
\begin{eqnarray}
\frac{d\hat{\cal C}}{d \tilde{t}} \sim A,
\end{eqnarray}
where $A$ is the proportional constant which may depend on $(\tilde{R}, d, M)$. From Fig.~1-3, one can see that $A$ is nearly independent of $\tilde{R}$ and $d$; While it strongly depends on $M$. By fitting the numerical data, it is found that $A \approx 0.4 M$.

From Fig.~1, one can also see that larger the subregion size $\tilde{R}$ is, longer time the linear growth stage lasts. It is expected that as $\tilde{R}$ approaches the cosmological horizon $\tilde{r} =1$ to cover the entire boundary space, the linear growth stage will last forever which agrees well with the CV conjecture. We can see this point more clearly in Fig.~4 where we take $d=2$ case as an example. We will give more evidences on this point later.

\begin{figure}[!htbp]
	\centering
	\includegraphics[width=0.5\textwidth]{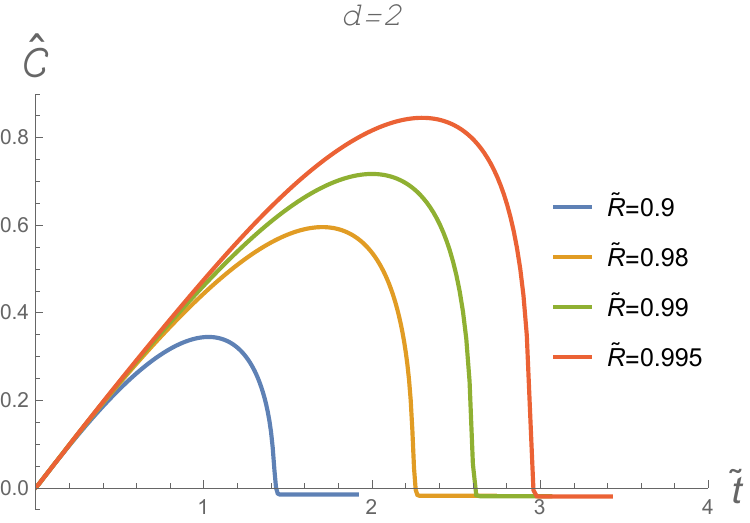}
	\caption{(colour online) Time evolution of normalised HSCV density $\hat{\cal C}$ for various $\tilde{R}$ with fixed $d$. The mass is fixed as $M=1$.}
\end{figure}

\subsubsection{Saturation time}

In Refs.~\cite{Fischler:2013fba,Zangeneh:2017tub}, one defines the saturation time as the time HEE approaches a constant. Similarly, for the complexity, we can also define a saturation time $\tilde{t}_{sat}$ as the time $\hat{\cal C}$ reaches its saturation constant $\hat{\cal C}_{sat}$.

In Fig.~5, we plot the time evolution of the two observables, $\hat{\cal S}$ and $\hat{\cal C}$ to make a comparison. From the figure, we can see the well-known fact that $\hat{\cal S}$ is always a monotonically increasing function of time. Moreover, from Fig.~5 and Fig.~2, one can see that $\hat{\cal S}$ and $\hat{\cal C}$ reache their saturation values at almost the same time. And their saturation time $\tilde{t}_{sat}$ is nearly independent of $d$ and $M$.

\begin{figure}[!htbp]
	\centering
	\includegraphics[width=0.45\textwidth]{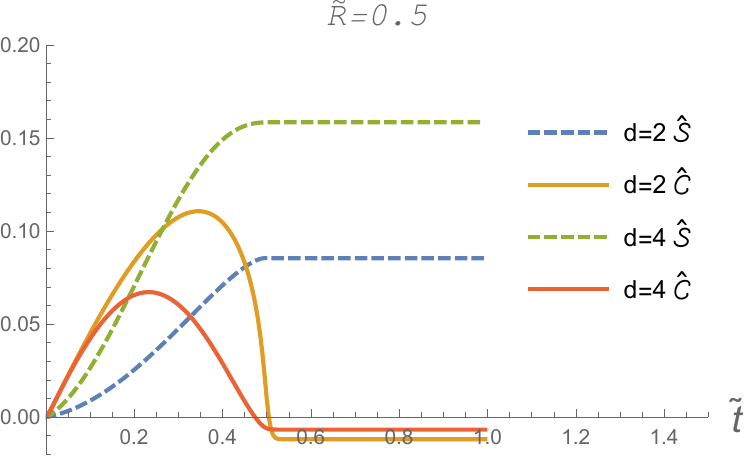}
	\includegraphics[width=0.45\textwidth]{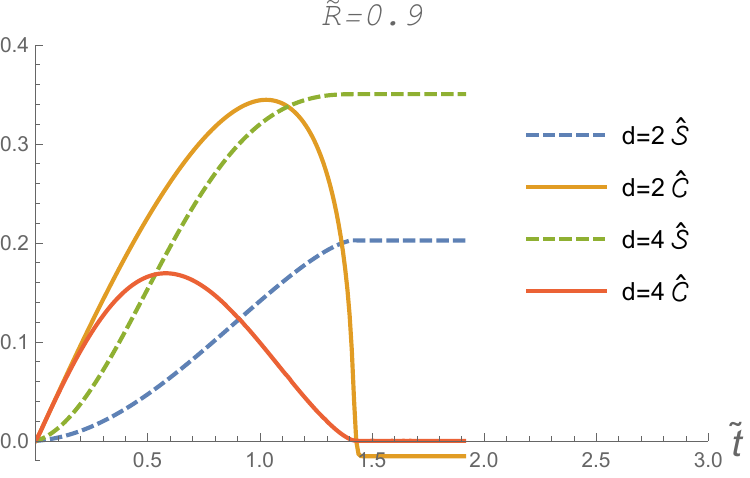}
	\caption{(colour online) Time evolution of HEE and $\hat{\cal C}$ for fixed $d$. The mass parameter is fixed as $M=1$.}
\end{figure}

In Fig.~6, the saturation time $\tilde{t}_{sat}$ for $\hat{\cal C}$  as a function of the subregion size $\tilde{R}$ is plotted. The numerical results can be well fitted by the function $\tilde{t}_{sat} = \tanh^{-1} (\tilde{R})$, as for the HEE \cite{Fischler:2013fba}. It is interesting to note that the $\tilde{t}_{sat}$ is just the time light takes travelling from the origin $\tilde{r}=0$ to the boundary of the subregion $\tilde{r} = \tilde{R}$.\footnote{We thank J.F. Pedraza for pointing out this.} From the figure and the fitting, one can easily see that $\tilde{t}_{sat}$ is linear in $\tilde{R}$ when $\tilde{R}$ is small; However, as $\tilde{R}$ approaches the cosmological horizon $\tilde{r} =1$, the saturation time diverges logarithmically and thus the linear growth stage will also last forever, as we already mentioned above.

\begin{figure}[!htbp]
	\centering
	\includegraphics[width=0.5\textwidth]{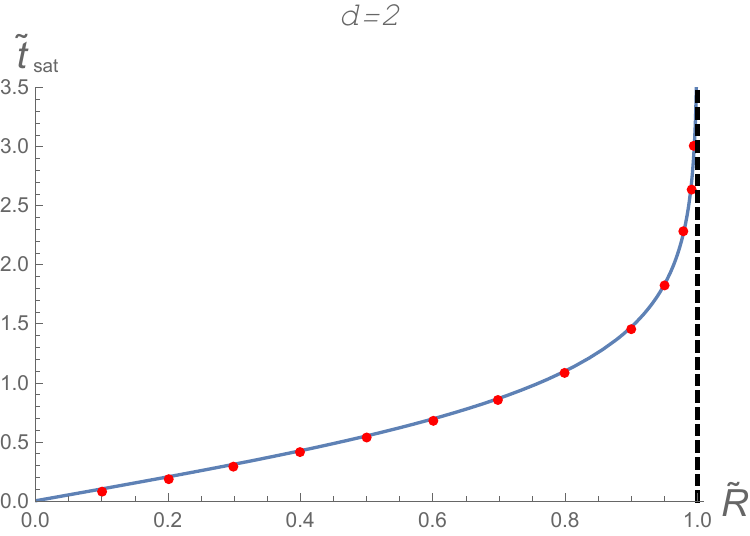}
	\caption{(colour online) Saturation time $\tilde{t}_{sat}$ as a function of the subregion size $\tilde{R}$. Red dots correspond to numerical results while the solid blue curve represents the fitting function $\tilde{t}_{sat} = \tanh^{-1} (\tilde{R})$. Space dimension and mass parameter are fixed as $d=2$ and $M=1$, respectively.}
\end{figure}

\section{Summary and Discussions}

In this work, we consider the holographic model of thermalization process for QFTs in dS spacetime. By applying the holographic subregion CV conjecture, we study the time evolution of subregion complexity under quench. The subregion ${\cal A}$ is chosen to be a sphere on the boundary time slice. The dual extremal codimension-one hypersurface $\Gamma_{{\cal A}}$ in the bulk, whose volume gives the complexity of ${\cal A}$, is found to be simply swept out by the HRT surface along the $\tilde{r}$-direction. The whole time evolution of subregion complexity can be divided into four stages: It first increases almost linearly; Then its growth slows down and after reaching a maximum it starts to drop down quickly, and shortly after the drop down stops and it gets to saturation finally. This picture is similar to that in flat boundary cases but with a strip subregion~\cite{Chapman:2018lsv,Ling:2018xpc}. This implies that the time evolution behaviours of subregion complexity are very general, and is independent of the subregion shape and the cosmological horizon.

The linear growth rate in the first stage is found to almost only depend on the mass parameter. As the subregion size approaches the cosmological horizon, this stage is expected to last forever, and as the HEE the saturation time is logarithmically divergent. The saturation time is found to depend almost only on the subregion size $\tilde{R}$, and their relation can be well fitted by the function $\tilde{t}_{sat} = \tanh^{-1} (\tilde{R})$. It is interesting to note that the $\tilde{t}_{sat}$ is just the time light takes travelling from the origin $\tilde{r}=0$ to the boundary of the subregion $\tilde{r} = \tilde{R}$. The underlying physical meaning of this fact needs further investigation.  

In this work, we only consider the HSCV conjecture. It is interesting to check whether general behaviours of subregion complexity still holds for other conjectures, for example the holographic subregion CA. In Ref.~\cite{Zhang:2014cga}, using HEE as a probe we show that including the Gauss-Bonnet correction will shorten the saturation time. It is also interesting to see how the higher-derivative terms affect the time evolution of subregion complexity. We leave these questions for further investigations.

\section*{Acknowledgement}

This work was supported by National Natural Science Foundation of China (Nos. 11605155 and 11675144). We thank J.F. Pedraza for helpful comments on this manuscript.

\end{document}